\documentclass[twocolumn,9pt]{article} 

\usepackage[square,numbers,sort&compress,comma]{natbib}

\usepackage{amsmath}
\usepackage{amssymb}
\usepackage{caption}
\usepackage{graphicx}
\usepackage{latexsym}
\usepackage{times}
\usepackage[pagewise]{lineno}
\usepackage{hyperref}

\topmargin - 12pt 
\renewenvironment{abstract}%
              {
               \small
               {\bfseries \abstractname}
               \par
               \vspace{10pt}
              }

\renewcommand\abstractname{Abstract}

\newcommand{\nomenclature}
              [1]
              {
               \bgroup
               \flushleft
               \small\bf
               #1
               \par
               \egroup
              }

\renewcommand{\section}
              [1]
              {
               \bgroup
               \flushleft
               \small\bf
               \refstepcounter{section}
               \arabic{section}. #1
               \par
               \egroup
              }

\renewcommand{\subsection}
              [1]
              {
               \bgroup
               \flushleft
               \small\em
               \refstepcounter{subsection}
               \arabic{section}.
               \arabic{subsection}. #1
               \par
               \egroup
              }

\renewcommand{\subsubsection}
              [1]
              {
               \bgroup
               \flushleft
               \small\em
               \refstepcounter{subsubsection}
               \arabic{section}.
               \arabic{subsection}.
               \arabic{subsubsection}. #1
               \par
               \egroup
              }

  \newcommand{\acknowledgement}
              [1]
              {
               \bgroup
               \flushleft
               \small\bf
               #1
               \par
               \egroup
              }

  \newcommand{\sectionbib}
              [1]
              {
               \bgroup
               \flushleft
               \small\bf
               #1
               \par
               \egroup
              }

\usepackage{bm}
\usepackage{physics}
\usepackage{xcolor}
\usepackage{tikz}
\usetikzlibrary{quantikz2}
\usepackage{adjustbox}

\graphicspath{{Figs/}}

\newcommand{\EQ}{\begin{equation}}
\newcommand{\EN}{\end{equation}}
\newcommand{\SEQ}{\begin{subequations}}
\newcommand{\SEN}{\end{subequations}}
\newcommand{\EQA}{\begin{eqnarray}}
\newcommand{\ENA}{\end{eqnarray}}
\newcommand{\CS}{\begin{dcases}}
\newcommand{\CN}{\end{dcases}}
\newcommand{\AR}{\begin{array}}
\newcommand{\AN}{\end{array}}

\newcommand{\mr}{\mathrm}

\newcommand{\bu}{\bm{u}}

\newcommand{\bA}{\bm{A}}
\newcommand{\bD}{\bm{D}}

\newcommand{\bI}{\bm{I}}
\newcommand{\bM}{\bm{M}}
\newcommand{\bS}{\bm{S}}
\newcommand{\bT}{\bm{T}}
\newcommand{\bV}{\bm{V}}
\newcommand{\bZ}{\bm{Z}}

\newcommand{\bphi}{\bm{\phi}}
\newcommand{\bvphi}{\bm{\varphi}}

\newcommand{\btheta}{\bm{\theta}}

\newcommand{\lrr}[1]{\left(#1\right)}
\newcommand{\lrs}[1]{\left[#1\right]}

\begin{document}



\small
\baselineskip 10pt

\setcounter{page}{1}
\title{\LARGE \bf Quantum computing of nonlinear reacting flows via the probability density function method}

\author{
{\large \noindent Jizhi Zhang$^{a,\#}$, Ziang Yang$^{a,\#}$, Zhaoyuan Meng$^{a,b}$, Zhen Lu$^{a,*}$, Yue Yang$^{a,c,*}$}
\\[10pt]
{\footnotesize \em $^a$State Key Laboratory for Turbulence and Complex Systems, School of Mechanics and Engineering Science,}\\[-5pt]
{\footnotesize \em Peking University, Beijing 100871, China}\\[-5pt]
{\footnotesize \em $^b$State Key Laboratory of Nonlinear Mechanics, Institute of Mechanics, Chinese Academy of Sciences, Beijing 100190, China}\\[-5pt]
{\footnotesize \em $^c$HEDPS-CAPT, Peking University, Beijing 100871, China}
}

\date{}  

\twocolumn[\begin{@twocolumnfalse}
\maketitle
\rule{\textwidth}{0.5pt}
\vspace{-5pt}

\begin{abstract} 
Quantum computing offers the promise of speedups for scientific computations, but its application to reacting flows is hindered by nonlinear source terms, the challenges of time-dependent simulations, and the difficulty of extracting meaningful physical quantities from quantum states. 
We employ a probability density function (PDF) formulation to transform the nonlinear reacting-flow governing equations into high-dimensional linear ones. 
The entire temporal evolution is then solved as a single large linear system using the history state method, which avoids the measurement bottleneck of conventional time-marching schemes and fully leverages the advantages of quantum linear system algorithms. 
To extract the quantity of interest from the resulting quantum state, we develop an efficient algorithm to measure the statistical moments of the PDF, bypassing the need for costly full-state tomography. 
A computational complexity analysis shows that the measurement algorithm achieves a complexity polynomial in the logarithm of the system size using low-order polynomial approximations, compared to the exponential cost of the exact operator, thereby retaining the quantum advantage gained from solving the linear system.
We validate the framework in two stages: an \textit{a priori} test confirms the accuracy of the measurement algorithm on beta distributions with known analytical moments, and a perfectly stirred reactor simulation demonstrates the capability to capture the PDF evolution and statistics of a nonlinear reactive system. 
This work establishes a pathway for applying quantum computing to nonlinear reacting flows.
\end{abstract}

\vspace{10pt}

{\bf Novelty and significance statement}

\vspace{10pt}

The simulation of reacting flows is limited by the prohibitive cost of resolving stiff, nonlinear chemical kinetics. 
Quantum computing offers an alternative paradigm, but its application is hindered by the inherent linearity of quantum mechanics and the difficulty of extracting physical information from quantum states.
This work addresses both challenges through a framework that integrates three components: a PDF formulation to linearize the governing equations; the history state method to solve the entire time-evolution in a single step, avoiding the measurement bottleneck; and an efficient algorithm to measure statistical moments from the final quantum state with polynomial complexity in the logarithm of the system size, bypassing exponentially costly full-state tomography.
This integration makes quantum simulation of nonlinear reacting flows viable, providing a pathway from problem formulation to physically meaningful output. 

\vspace{5pt}
\parbox{1.0\textwidth}{\footnotesize {\em Keywords:} 
Quantum computing; 
Reacting flows;
Probability density function; 
Statistics measurement}
\rule{\textwidth}{0.5pt}
*Corresponding author: Zhen Lu (zhen.lu@pku.edu.cn) and Yue Yang (yyg@pku.edu.cn)
\vspace{5pt}
\end{@twocolumnfalse}] 

\section{Introduction\label{sec:intro}} \addvspace{10pt}

The simulation of reacting flows faces a formidable obstacle rooted in the chemical source terms, which are highly nonlinear and exhibit stiffness across a wide range of temporal scales~\cite{Masri2021,Pope2013}. 
The computational cost of resolving these nonlinear dynamics scales unfavorably with the number of species and the complexity of the reaction mechanism, creating a significant bottleneck for high-fidelity simulations~\cite{Pope2013,Ren2014,Domingo2023}. 
This challenge has motivated a sustained search for new computational strategies~\cite{Zhou2022,Uranakara2023Accelerating,Ihme2024,Zhang2022,Mao2023,Deng2025,Koenig2025}.

Quantum computing offers a new paradigm for scientific computation, with provable speedups for certain classes of problems~\cite{Nielsen2010,Givi2020,Givi2021}. 
Current quantum computer, however, remains in the noisy intermediate-scale quantum (NISQ) era~\cite{Preskill2018}, where limited qubit counts and high error rates~\cite{Yang2026} preclude the execution of large-scale, fault-tolerant algorithms~\cite{Hoefler2023}. 
Nevertheless, the development of quantum algorithms targeting the fault-tolerant era is essential to establish theoretical foundations and identify the problems where quantum advantage is most likely to be realized~\cite{Meng2025}. 
A fundamental obstacle is that quantum mechanics is governed by linear evolution, making quantum computers natively ill-suited for the nonlinear dynamics~\cite{Tennie2025}.
Bridging this gap between the linearity of quantum computation and the nonlinearity of reacting flows is a prerequisite for any quantum approach to combustion simulation.

Several linearization techniques have been explored for quantum computing of nonlinear dynamics~\cite{Liu2021,Akiba2023,Akiba2025,Sanavio2024,Joseph2020,Novikau2025,Zhang2025,Xue2025,Meng2023,Meng2024a,Meng2024b,Succi2024Ensemble,Jin2023,Lee2026,Gonzalez-Conde2025}.
The Carleman linearization lifts the state space to higher-order tensor products to embed a nonlinear system into a linear one~\cite{Liu2021,Akiba2023,Akiba2025,Sanavio2024}. 
The Koopman–von Neumann formulation achieves linearization by evolving observable functions~\cite{Joseph2020,Novikau2025,Zhang2025}. 
Other approaches include the quantum homotopy analysis method, which iteratively linearizes the governing equations~\cite{Xue2025}, and the hydrodynamic Schr\"{o}dinger equation~\cite{Meng2023,Meng2024a,Meng2024b}, which recasts fluid dynamics in a Schr\"{o}dinger-Pauli equation. 
These methods have advanced the field, but each introduces specific trade-offs in accuracy or applicability. 
The Carleman linearization requires truncation of an infinite hierarchy and can suffer from convergence issues for strongly nonlinear systems~\cite{Liu2021}. 
The Koopman-based methods face challenges in identifying suitable observable spaces~\cite{Zhang2025}. 
For combustion applications, a formulation that can handle the nonlinear chemical source term without truncation or approximation of the nonlinearity itself would be particularly attractive.

The transported probability density function (PDF) method provides precisely such a formulation~\cite{Pope1985,Haworth2010,Gourianov2025}.
In the PDF framework, the nonlinear chemical source term appears as a linear transport operator in composition space~\cite{Pope1985}, making the governing equation a high-dimensional Fokker–Planck-type equation amenable to quantum algorithms~\cite{Harrow2009,Costa2022,Wang2024,An2023,Jin2024,Lu2024,Brearley2024,Tennie2024,Over2025}. 
The connection between the PDF method and quantum computation was first explored by Xu et al.~\cite{Xu2018,Xu2019}, who computed the reactant conversion rate via Monte Carlo simulation of stochastic particles. 
In contrast, the present work solves the PDF transport equation directly as a linear PDE, which maps naturally onto the matrix formalism of quantum linear system algorithms (QLSA).
Separately, Lu and Yang~\cite{Lu2024} demonstrated quantum computing of reacting flows via Hamiltonian simulation, which handles time evolution naturally but does not address the nonlinearity of the chemical source term. 
The PDF formulation adopted here eliminates the nonlinearity of source terms at the cost of increasing the problem dimensionality, a trade-off that quantum computing is uniquely positioned to manage given the exponential scaling of the quantum state space~\cite{Nielsen2010}.

Despite the linearization, two significant challenges remain for a practical quantum simulation framework. 
First, the PDF transport equation is time-dependent, and a naive application of QLSA would require iterative calls at each time step, with intermediate measurements to extract the system state before proceeding~\cite{Liu2023Quantum,Ye2024hybrid,Bharadwaj2025Compact,Song2025Incompressible}. 
This measurement bottleneck severely diminishes the potential for quantum speedup. 
Second, even after the quantum state encoding the PDF is obtained, extracting meaningful physical quantities from it is non-trivial~\cite{Su2025}. 
The statistics of interest, such as the mean and variance, are encoded in the amplitudes of the quantum state, but full quantum state tomography to reconstruct these amplitudes is itself exponentially costly~\cite{Nielsen2010,Aaronson2015}. 
An efficient method for measuring the relevant statistics from the quantum state is therefore essential to close the gap between solving the linear system and obtaining physically useful results.

In this work, we present a quantum computing framework that addresses both challenges to simulate nonlinear reacting flows.
We employ the history state method~\cite{Berry2014,Jin2023} to formulate the temporal evolution of the PDF transport equation as a single, large linear system solvable with one QLSA call, thereby bypassing the measurement bottleneck of iterative time-marching schemes.
To address the information extraction problem, we develop an efficient algorithm for measuring statistical moments of the PDF from the final quantum state. 
This algorithm exploits low-order polynomial approximations of the measurement operator to achieve a complexity polynomial in $\log N$, where $N$ is the system size, without requiring full state tomography.
We validate the framework by simulating a perfectly stirred reactor, demonstrating its capability to capture the PDF evolution and statistics of a nonlinear reactive system. 
Together, these elements establish a complete and viable pathway for applying quantum computing to tackle challenges in the simulation of nonlinear reacting flows.

The remainder of this paper is organized as follows. 
Section~\ref{sec:PDF} describes the PDF transport equation and the history state formulation. 
Section~\ref{sec:stats} presents the statistics measurement algorithm and its complexity analysis. 
Section~\ref{sec:results} provides the validation results, and Sec.~\ref{sec:concls} concludes the paper.

\section{PDF evolution\label{sec:PDF}} \addvspace{10pt}

\begin{figure*}[!ht]
\centering
\vspace{-0.4 in}
\includegraphics[width=\linewidth]{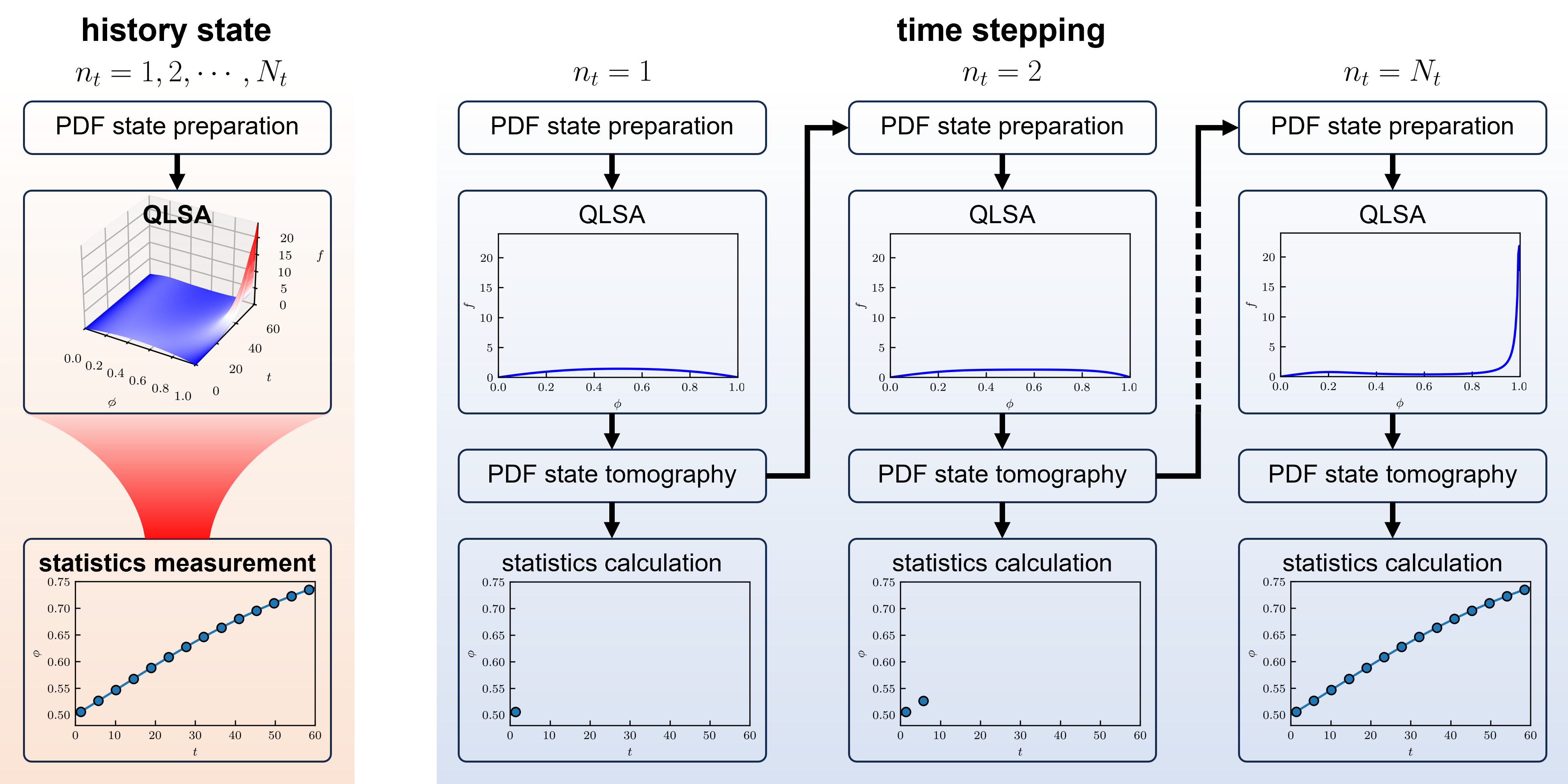}
\vspace{10 pt}
\caption{\footnotesize 
Comparison of our history state method (left) with a conventional time-stepping approach (right) for solving the PDF transport equation. 
Our history state method solves the entire space-time evolution with a single QLSA call, avoiding the measurement bottleneck in the iterative conventional approach.
}
\label{fig:approach}
\end{figure*}

The compositions $\bphi$ in reactive flow evolve as $\mr{D} \bphi / \mr{D} t = \bM + \bS\lrr{\bphi}$, where $t$ is the time, and $\bM$ and $\bS$ are the mixing and reaction terms, respectively. 
The nonlinear reaction term $\bS\lrr{\bphi}$ poses a significant challenge for quantum algorithms.
To overcome this, we adopt the transported PDF method~\cite{Pope1985,Haworth2010}. 
It recasts the problem in terms of the one-point, one-time Eulerian PDF, $f_{\bphi}\lrr{\bvphi}$, transported as
\EQ\label{eq:PDF}
\pdv{f_{\bphi}}{t} 
\!+\! 
\overline{\bu\vert\bvphi}\cdot\nabla f_{\bphi} 
\!=\!
-\!\nabla_{\bvphi}\cdot\lrs{f_{\bphi}\lrr{\overline{\bM\vert\bvphi}\!+\!\bS}},
\EN
where $\bu$ is the velocity, $\bvphi$ is the sample-space variable, and $\overline{\cdot\vert\bvphi}$ denotes the conditional mean upon $\bphi=\bvphi$. 
The central advantage of this formulation is that the nonlinear reaction term $\bS$ appears in closed form as a linear transport operator in composition space, requiring no modeling. 
For a system of $n_c$ compositions, the PDF is defined over the $n_c$-dimensional composition space, making Eq.~\eqref{eq:PDF} a high-dimensional Fokker-Planck-type equation.
This high dimensionality, which poses a major difficulty for classical numerical methods, is naturally suited to quantum computation.

We discretize the PDF transport equation in Eq.~\eqref{eq:PDF} into a system of linear algebraic equations. 
A central difference scheme is employed for the spatial derivatives in both physical and composition spaces, and implicit Euler integration is used for temporal advancement. 
At each time step, the discretized system is expressed as $\bm{T}\bm{f}_\phi^{k+1}=\bm{f}_\phi^k$, where $\bm{f}_\phi^k$ denotes the vector of discrete PDF values at time step $k$, and the matrix $\bm{T}$ is determined by the discretization over $N_\phi$ grid points.

Rather than solving this linear system iteratively at each time step, we employ the history state method~\cite{Berry2014,Jin2023} to couple all time steps into a single global linear system.
By stacking the solution vectors from all $N_t$ time steps, the entire temporal evolution is formulated as
\EQ\label{eq:hs}
    \bA \ket{\psi} = \ket{b},
\EN
where the matrix $\bA$ has a block bidiagonal structure with $\bT$ on the diagonal and $-\bI$ on the subdiagonal, $\ket{\psi}$ denotes the final quantum state encoding the complete PDF evolution, and the right-hand side $\ket{b}$ encodes the initial PDF distribution in its first block. 
The resulting matrix $\bA$ inherits the sparsity of $\bT$,  which is essential for efficient implementation of QLSA. 
The total system size is $N = N_\phi\times N_t$, requiring $ n = \log(N_\phi) + \log(N_t) $ qubits to encode $\ket{\psi}$ and $\ket{b}$. 
This formulation allows the QLSA to be applied only once, avoiding the intermediate measurements required by conventional time-marching approaches thereby fully leverages the potential exponential speedup of QLSA.

\section{Statistics measurement\label{sec:stats}} \addvspace{10pt}

\subsection{Measurement algorithm} \addvspace{10pt}

Full reconstruction of the PDF from the quantum state $\ket{\psi}$ via tomography is prohibitively expensive. 
Instead, we typically seek statistics $\overline{q\lrr{\bvphi}} = \int_{-\infty}^{\infty} q\lrr{\bvphi} f_{\bphi}\lrr{\bvphi} d\bvphi$, where $q\lrr{\bvphi}$ is a function of $\bvphi$.
For simplicity, we consider a single composition variable $\phi$ normalized to the range $\lrs{0, 1}$, as is common for reactive scalars.
This moment can be obtained through quantum phase estimation of an observable $\dyad{q}$~\cite{Xu2018,Xu2019} or measuring the inner product $\braket{q}{\psi}$~\cite{Jin2023,Xu2026}, where $\ket{q}$ encodes the function $q\lrr{\bvphi}$.
However, preparing an arbitrary quantum state $\ket{q}$ is non-trivial, requiring an optimal circuit depth of $\mathcal{O}\lrr{2^n/n}$~\cite{Sun2023_Asymptotically}.
Consequently, this resource overhead would diminish any quantum advantage gained from solving the linear PDEs.

We propose an efficient method to measure statistics from $\ket{\psi}$ via the Hadamard test.
We define a unitary operator
\EQ\label{eq:U}
    U = \sum_{j=0}^{N-1} e^{i\theta_j}\dyad{j},
\EN
where $i=\sqrt{-1}$, $\theta_j = \arccos{q\lrr{\varphi_j}}$, and $\ket{j}$ is the computational basis, with $j=0, 1, \ldots, N-1$. 
The state $UH^{\otimes n}\ket{0}^{\otimes n}$ encodes the function $q\lrr{\bvphi}$ into the amplitudes of a uniform superposition, since $H^{\otimes n}\ket{0}^{\otimes n}$ produces an equal superposition over all basis states and $U$ imprints $\cos{\theta_j}=q\lrr{\varphi_j}$ onto each component, where $H$ is the Hadamard gate.

The circuit in Fig.~\ref{fig:circuit} operates as follows.
First, a Hadamard gate $H$ places the ancilla qubit in an equal superposition $\lrr{\ket{0}+\ket{1}}/\sqrt{2}$. 
A controlled $U_\psi$ gate then loads the PDF solution $\ket{\psi}$ onto the register, conditioned on the ancilla being in state $\ket{1}$. 
The Pauli-X gate flips the ancilla, after which a controlled $UH^{\otimes n}$ loads the state encoding $q\lrr{\bvphi}$, conditioned on the ancilla now being $\ket{1}$. 
The resulting state is $\lrr{\ket{0}_a\ket{\psi}+\ket{1}_aUH^{\otimes n}\ket{0}^{\otimes n}}/\sqrt{2}$.
A final Hadamard gate on the ancilla creates interference between the two loaded states~\cite{Cleve1998}, producing 
$\frac{1}{2}\ket{0}_a\otimes\lrr{\ket{\psi}+UH^{\otimes n}\ket{0}^{\otimes n}}+\frac{1}{2}\ket{1}_a\otimes\lrr{\ket{\psi}-UH^{\otimes n}\ket{0}^{\otimes n}}$. 
Measuring the Pauli-Z expectation on the ancilla gives $\expval{Z_a}=\mathrm{Re}\bra{\psi}UH^{\otimes n}\ket{0}^{\otimes n}$. 
Since $UH^{\otimes n}\ket{0}^{\otimes n}=\sum_{j=0}^{N-1} e^{i\theta_j}\ket{j}$, this evaluates to
\EQ\label{eq:exp}
    \expval{Z_a} = \dfrac{1}{\sqrt{N}}\sum_{j=0}^{N-1} \psi_j q\lrr{\varphi_j}
\EN
by the definition $\theta_j=\arccos{q\lrr{\varphi_j}}$, where $\psi_j$ denotes the amplitude of $\ket{\psi}$ in the computational basis. 
With the PDF encoded as $\psi_j \propto f_\phi\lrr{\varphi_j}$ and the grid spacing absorbed into the normalization, Eq.~\eqref{eq:exp} yields the desired statistics. 
Decomposition and effective approximation of $U$ into Pauli operators are introduced in the next subsection.

\begin{figure}[ht!]
\centering
\begin{adjustbox}{width=192pt}
\begin{quantikz}[row sep=1em, column sep=1em]
    \lstick{$\ket{0}_a$} & \gate{H} & \ctrl{1} & \gate{X} & \ctrl{1} & \gate{H} & \meter{} \\
    \lstick{$\ket{0}^{\otimes n}$} & \qw & \gate{U_\psi} & \qw & \gate{UH^{\otimes n}} & \qw & \qw
\end{quantikz}
\end{adjustbox}
\caption{\footnotesize Quantum circuit to measure the statistics from $\ket{\psi}$ encoding PDF on $n$ qubits. The ancilla qubit controls the loading of the PDF state $\ket{\psi}$ via $U_\psi$ and the measurement function state via $UH^{\otimes n}$. Interference through the Hadamard gates encodes the desired statistic into the ancilla, which is then obtained by measuring the Pauli-Z expectation value.}
\label{fig:circuit}
\end{figure}

\subsection{Decomposition and approximation of $U$} \addvspace{10pt}

To implement the quantum circuit in Fig.~\ref{fig:circuit}, we decompose the operator $U$ into quantum gates. 
The unitary transform in Eq.~\eqref{eq:U} is re-expressed as
\EQ\label{eq:Udecomp}
    U = \exp\lrr{i\sum_{j=0}^{N-1}\eta_j \bD^j},
\EN
where the coefficients $\eta_j$ are found by solving the linear system $\bV\lrr{0, 1, \cdots, N-1}\bm{\eta} = \btheta$, with the Vandermonde matrix $\bV$.
Here, the diagonal matrix $\bD=\mr{diag}\lrr{0, 1, \cdots, 2^n-1}$ is decomposed~\cite{Meng2023,Meng2024b} in terms of Pauli operators as
\EQ\label{eq:Ddecomp}
    \bD = \dfrac12\lrs{\lrr{2^n-1}I_2^{\otimes n}-\sum_{j=1}^n 2^{n-j} \hat{\bZ}_j},
\EN
where $I_2$ and $Z$ are the unit and Pauli-Z operators, respectively, and $\hat{\bZ}_j = I_2^{\otimes j-1}\otimes Z\otimes I_2^{\otimes n-j}$.

Alternatively, $U$ can be approximated using an $m$-th order polynomial, 
\EQ\label{eq:Uapprox}
    U \approx \exp \lrr{ i \sum_{j=0}^m \xi_j \bD^j },
\EN
where the coefficients $\xi_j$ are obtained via polynomial fitting of the phase function $\theta_j$.
This approximation is effective because the phase functions for quantities of physical interest are smooth and well-approximated by low-order polynomials. 
Furthermore, the exact decomposition in Eq.~\eqref{eq:Udecomp} guarantees convergence, as the polynomial interpolation becomes exact for order $m=N-1$.

\begin{figure}[!ht]
    \centering
    \includegraphics[width=192pt]{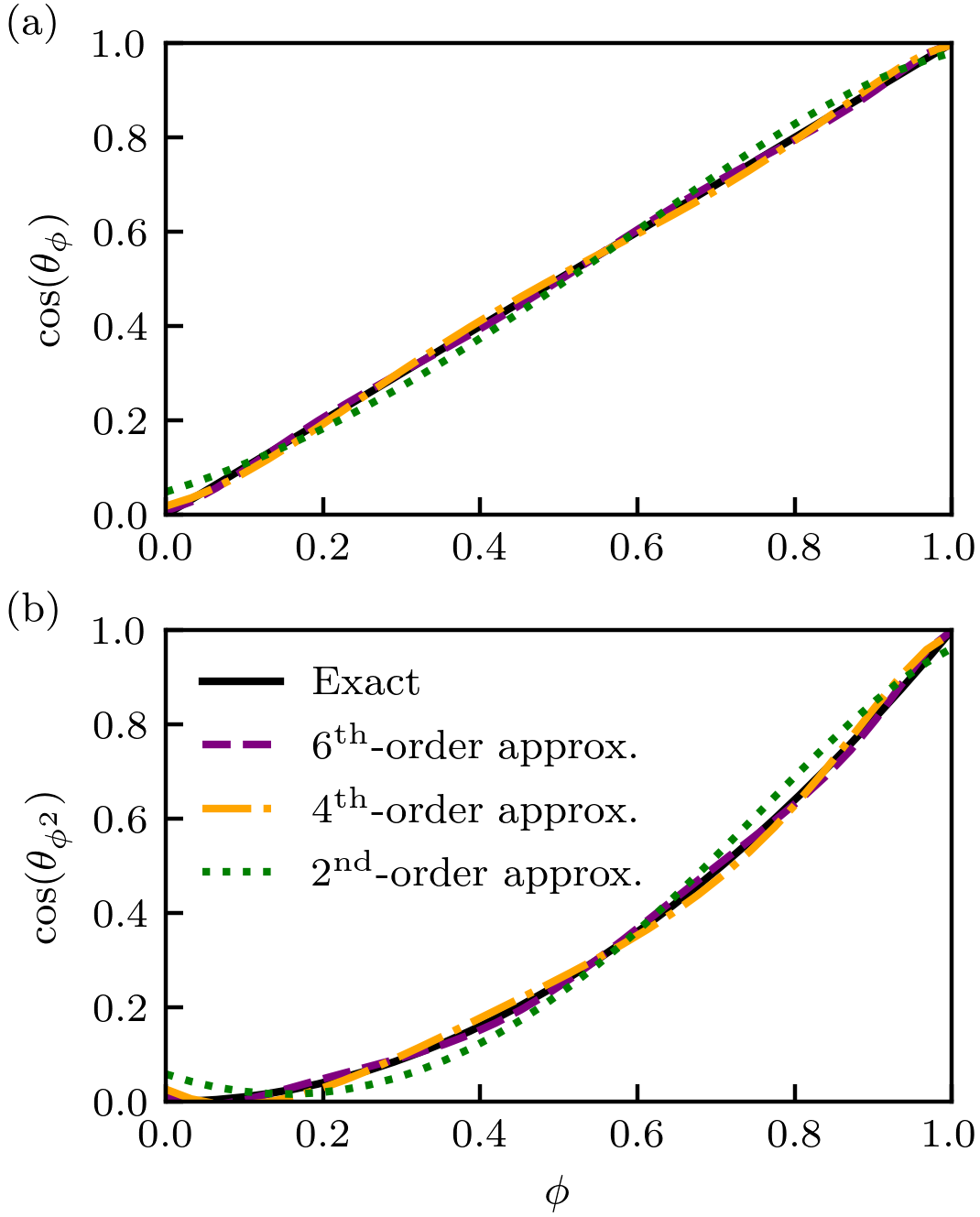}
    \caption{Accuracy of the polynomial approximation of the measurement operator for (a) the mean, $\cos{\theta_\phi}$, and (b) the variance, $\cos{\theta_{\phi^2}}$. The exact phase functions (solid black lines) are compared with second- (dotted green), fourth- (dash-dotted orange), and sixth-order (dashed purple) polynomial approximations.}
    \label{fig:coeff}
\end{figure}

Figure~\ref{fig:coeff} illustrates the accuracy of the polynomial approximation by comparing $\cos{\theta_\phi}$ and $\cos{\theta_{\phi^2}}$ for mean and variance obtained from the exact operator and its second-, fourth-, and sixth-order approximations. 
For the mean, the second-order approximation already captures the overall trend, with deviations most visible in the intermediate range. 
For the variance, the greater nonlinearity of $\cos{\theta_{\phi^2}}$  leads to larger deviations at low order, particularly for small $\phi$ values. 
In both cases, the sixth-order approximation is nearly indistinguishable from the exact result across the domain, confirming that low-order polynomials provide an accurate and efficient representation of the measurement operator.


\subsection{Complexity analysis} \addvspace{10pt}

\begin{figure}[ht!]
    \centering
    \includegraphics[width=192pt]{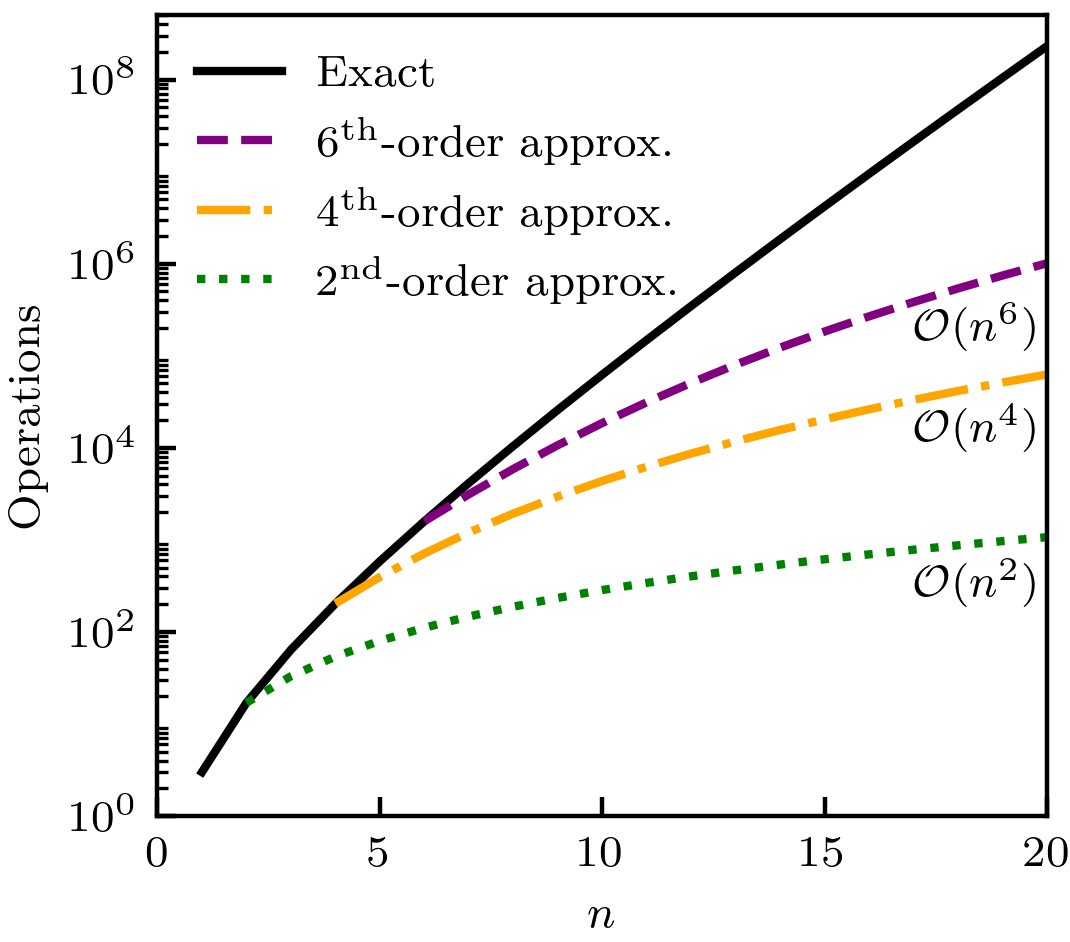}
    \caption{Operational cost (number of CNOT and $R_z$ gates) for the exact operator $U$ and its second-, fourth-, and sixth-order polynomial approximations as a function of the number of qubits $n$.}
    \label{fig:complexity}
\end{figure}

Implementing the $k$-th order term $\exp\lrr{i\eta_k \bD^k}$ in Eq.~\eqref{eq:Udecomp} requires $ \sum_{j=1}^k C_k^j \lrr{2j+1} $ CNOT and $R_z$ gates.
The exact decomposition of Eq.~\eqref{eq:Udecomp} thus allows for measuring an arbitrary statistic of $q$ in $2^{n-1}\lrr{n^2+2n+2} - (n+1)$ operations, a complexity of $\mathcal{O}\lrr{n^2 2^{n-1}}$. 
This cost is comparable to $\mathcal{O}\lrr{2^n/n}$ required for arbitrary $\ket{q}$~\cite{Sun2023_Asymptotically}, and exceeds the classical $\mathcal{O}\lrr{2^n}$ calculation, which would diminish the quantum advantage gained from solving the linear system via QLSA.
In contrast, the $m$-th order approximation in Eq.~\eqref{eq:Uapprox} corresponds to a complexity of $\mathcal{O}\lrr{n^m}$, which is polynomial in $n=\log N$. 
As demonstrated in Fig.~\ref{fig:coeff}, low-order approximations provide sufficient accuracy for statistics such as the mean and variance, making efficient measurement practical. 
Figure~\ref{fig:complexity} compares the operational cost for the exact operator $U$ and its low-order approximations. 
The polynomial scaling of the approximate scheme stands in stark contrast to the exponential growth of both the exact decomposition and classical computation, confirming that the proposed measurement algorithm preserves the quantum advantage gained from solving the linear system via QLSA.

\section{Validations\label{sec:results}} \addvspace{10pt}

We validate the proposed framework in two stages. 
First, the statistics measurement algorithm developed in Sec.~\ref{sec:stats} is validated \textit{a priori} by applying it to beta distributions with known analytical moments, independent of the QLSA solution. 
This isolates and verifies the accuracy of the measurement scheme itself. 
Second, the complete framework is demonstrated on a perfectly stirred reactor (PSR) model, where we simulate the PDF evolution via the history state method and extract the statistical moments from the resulting quantum state. 

Note that the current quantum computer is in the NISQ era~\cite{Preskill2018,Hoefler2023}. 
In this regime, imperfect qubit control introduces noise, while the number and connectivity of available qubits remain limited.
Consequently, following Ref.~\cite{Lu2024}, the simulations are conducted using a classical statevector simulator, yielding exact quantum state amplitudes and expectation values in the absence of quantum noise and sampling errors.

\subsection{\textit{a priori} validation of the measurement algorithm\label{sec:apriori}} \addvspace{10pt}

To validate the measurement algorithm independently, we apply it to beta distributions with known analytical moments. 
The beta distribution is parameterized by $\alpha$ and $\beta$, for which the mean and variance are available. 
The distributions are discretized on $n=10$ qubits, and the mean and variance are computed using the second-, fourth-, and sixth-order polynomial approximations of the measurement operator $U$.

\begin{figure*}
    \centering
    \includegraphics[width=\linewidth]{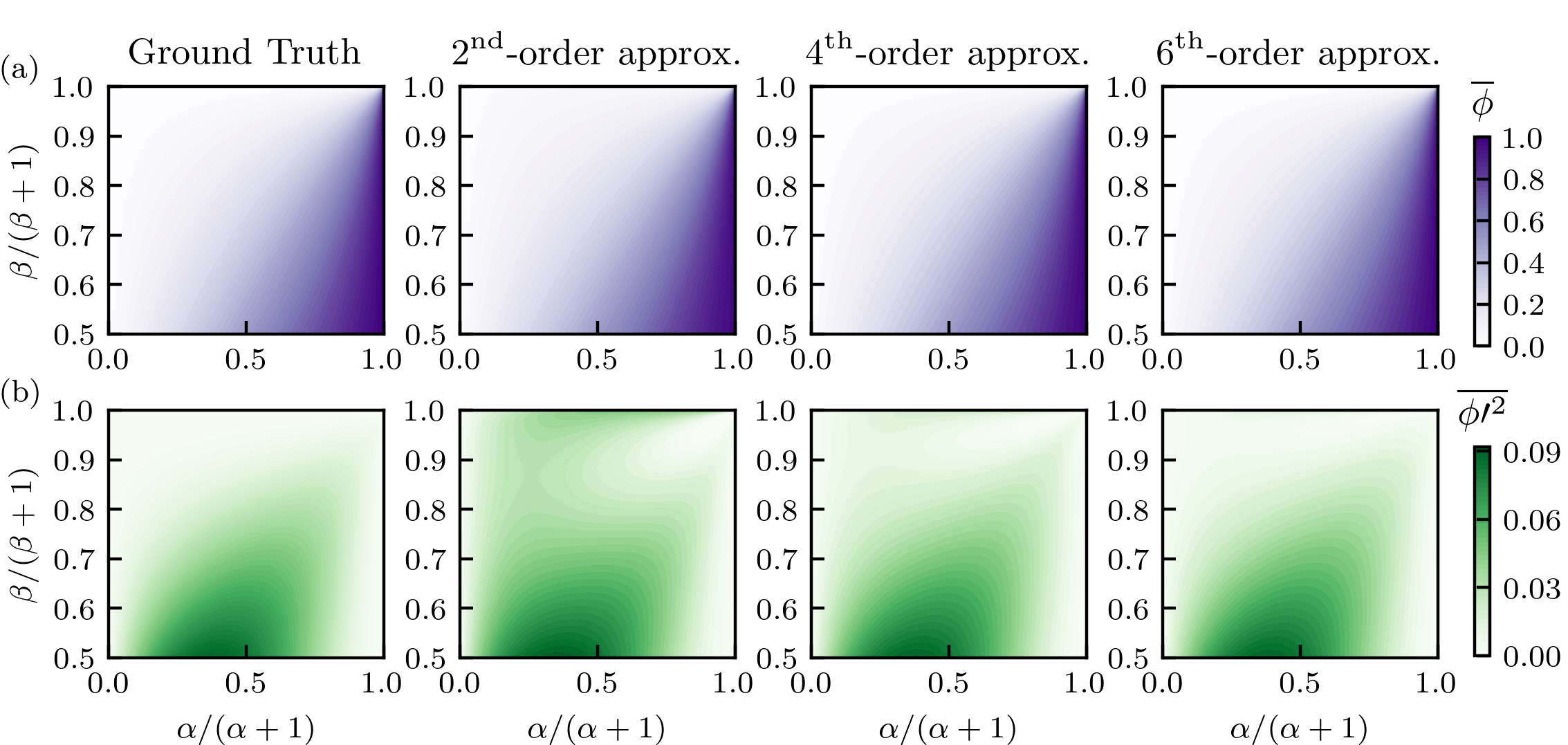}
    \caption{\textit{A priori} validation of the statistics measurement algorithm on beta distributions with parameters $\alpha$ and $\beta$ on (a) mean $\overline{\phi}$ and (b) variance $\overline{\phi\prime^2}$. The ground-truth analytical values (leftmost panels) are compared with second-, fourth-, and sixth-order polynomial approximations of the measurement operator. The axes are normalized as $\alpha/(\alpha+1)$ and $\beta/(\beta+1)$.}
    \label{fig:exp}
\end{figure*}

Figure~\ref{fig:exp} compares the ground-truth statistics against the values obtained from the low-order approximations across the $\lrr{\alpha, \beta}$ parameter space. 
For the mean, even the second-order approximation reproduces the analytical values with good fidelity across the entire parameter range, consistent with the near-linear behavior of the phase function $\cos{\theta_\phi}$ observed in Fig.~\ref{fig:coeff}(a). 
For the variance, the second-order approximation exhibits visible discrepancies. 
This is expected from the greater nonlinearity of $\cos{\theta_{\phi^2}}$ shown in Fig.~\ref{fig:coeff}(b), which demands higher-order polynomials for accurate representation. 
The fourth-order approximation substantially reduces these errors, and the sixth-order result is visually indistinguishable from the ground truth for both statistics. 
These results confirm that the proposed measurement algorithm accurately extracts statistical moments from quantum states encoding a wide variety of PDF shapes, and that low-order polynomial approximations are sufficient for practical accuracy.

\subsection{PDF evolution} \addvspace{10pt}

We demonstrate the complete framework on a PSR with a single reactive scalar $\phi$. 
The reaction $\bS\lrr{\phi}$ and mixing $\bM$ terms in Eq.~\eqref{eq:PDF} are modeled as~\cite{Lu2017Analysis}
\EQ
S\!=\!15\lrr{1\!-\!\phi}\exp\lrr{\!-\dfrac{\phi_a}{\phi+\phi_i}\!}
\mr{and}\; 
M\!=\!-\dfrac{\phi}{4}.
\EN
The parameters $\phi_i$ and $\phi_a$ are set as 0.15 and 1.8, respectively. 
By restricting the problem to zero spatial dimensions, this model isolates the challenges of nonlinear chemical kinetics while preserving the essential characteristics of reacting systems.
The PDF transport equation is solved using the history state method with the Harrow–Hassidim–Lloyd (HHL) algorithm~\cite{Harrow2009} as the QLSA,  implemented on a quantum simulator. 
The composition and temporal spaces are each discretized using 5 and 3 qubits, respectively, due to the ancilla qubit overhead of HHL and the memory limitations of the classical workstation.
For brevity, one representative case is shown below.
The initial PDF is a beta distribution with $\alpha=2$ and $\beta=4$. 

\begin{figure}[ht!]
    \centering
    \includegraphics[width=192pt]{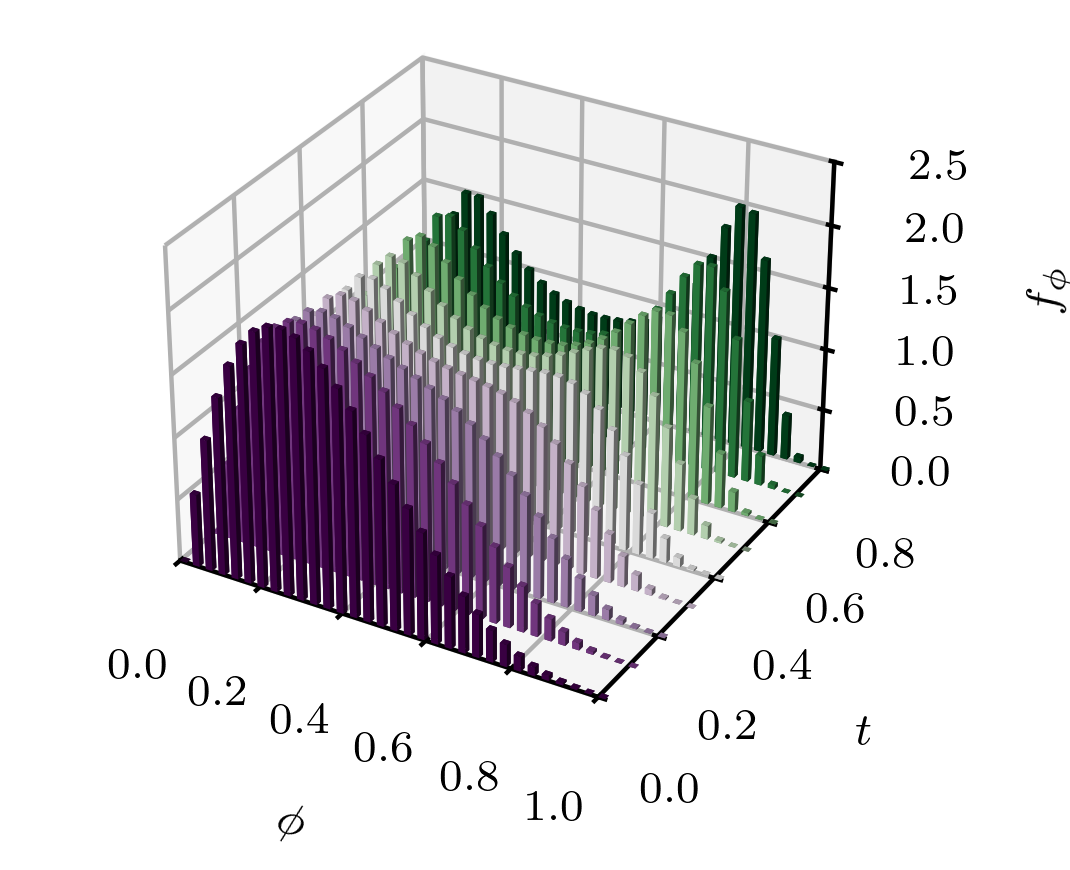}
    \caption{PDF evolution obtained via the history state method, using 5 and 3 qubits ($N_\phi=32$ and $N_t=8$) for compositional and temporal spaces, respectively.}
    \label{fig:pdf}
\end{figure}

Figure~\ref{fig:pdf} shows the evolution of the scalar PDF over time, obtained from the history state $\ket{\psi}$. 
Starting from an initial beta distribution centered around $\phi=0.5$, the PDF is driven by reaction and mixing. 
The distribution first broadens and shifts towards higher $\phi$ values due to reaction, then evolves towards a bimodal shape, before finally collapsing into a sharp peak at the steady-state value near $\phi=0.8$. 
This complete temporal evolution, capturing the transient dynamics from a smooth initial distribution to a sharp steady state, is obtained from a single quantum computation via the history state method. 
The results obtained from the quantum simulator are in agreement with those from a classical finite-difference solution of the same discretized system, confirming the correctness of the history state formulation. 
This demonstrates the framework's ability to resolve the nonlinear dynamics of a reactive system within a quantum computing workflow.

\subsection{Statistics measurement} \addvspace{10pt}

\begin{figure}[ht!]
    \centering
    \includegraphics[width=192pt]{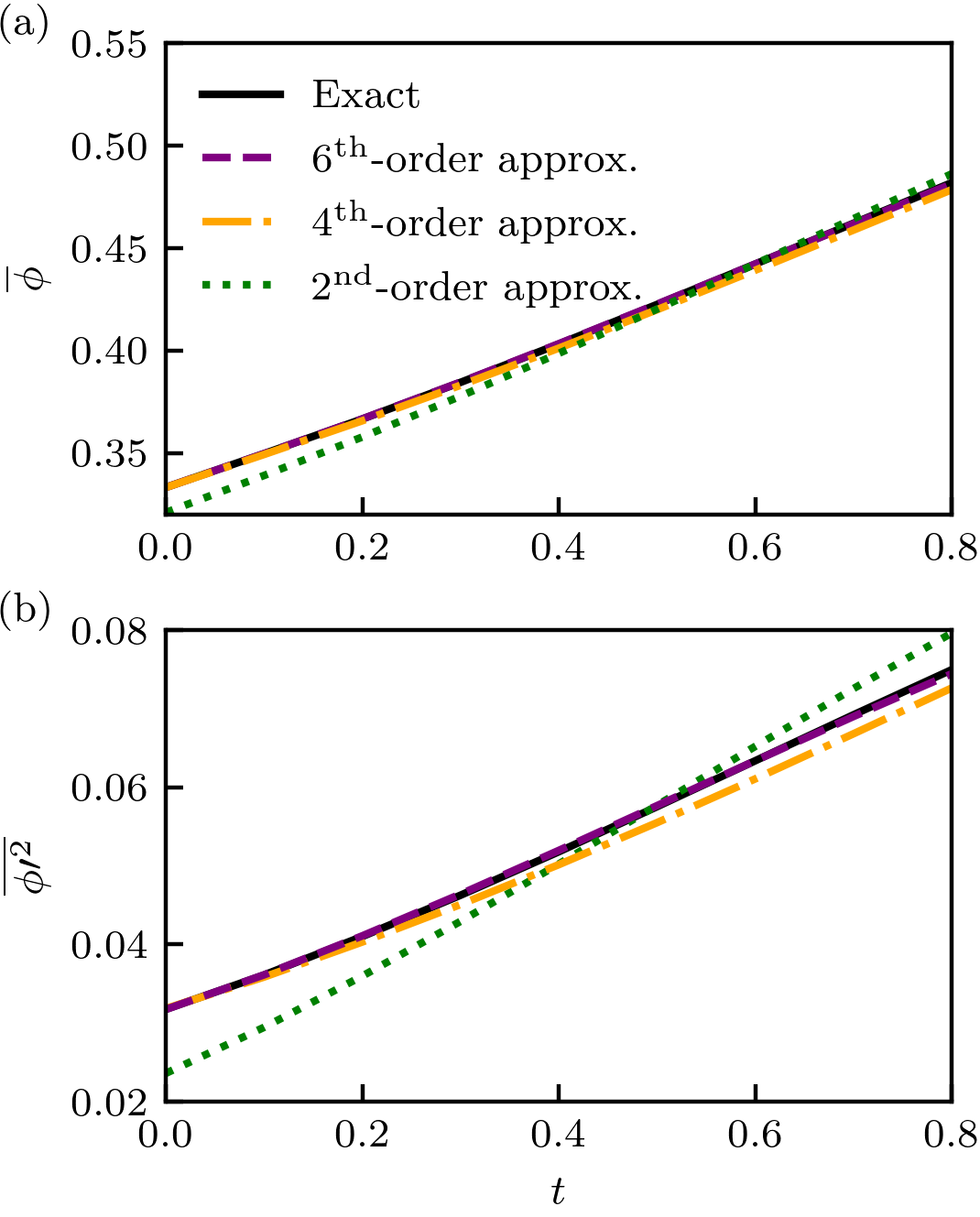}
    \caption{Mean $\overline{\phi}$ and variance $\overline{\phi'^2}$ obtained with exact measurement operator $U$ and its low-order approximations.}
    \label{fig:stats}
\end{figure}

We extract the mean $\overline{\phi}$ and variance $\overline{\phi'^2}$ from the quantum state $\ket{\psi}$ using the measurement circuit in Fig.~\ref{fig:circuit}, with the exact operator $U$ and its second-, fourth-, and sixth-order polynomial approximations. 
Figure~\ref{fig:stats} compares the time evolution of both statistics obtained from each approximation against the exact measurement. 
All approximations capture the temporal trends of both the mean and variance. 
The averaged relative errors are 1.55\%, 0.44\%, and 0.03\% on $\overline{\phi}$ and 8.98\%, 2.47\%, and 0.24\% on $\overline{\phi'^2}$ for the second-, fourth-, and sixth-order approximations, respectively. 
Consistent with the \textit{a priori} results in Sec.~\ref{sec:apriori} and Fig.~\ref{fig:coeff},  the mean converges more rapidly than the variance due to the simpler structure of its phase function. 
As the polynomial order increases, the approximate moments converge rapidly to the exact values, confirming that the low-order polynomial approximations provide both the accuracy and the polynomial computational complexity needed for practical statistics extraction from the quantum state. 

\section{Conclusions\label{sec:concls}} \addvspace{10pt}

We presented a quantum computing framework for simulating nonlinear reacting flows. 
First, a PDF formulation transforms the nonlinear governing equations into a high-dimensional linear Fokker–Planck-type equation, making the problem amenable to quantum linear solvers. 
Second, the history state method encodes the entire temporal evolution into a single linear system solvable with one QLSA call, bypassing the measurement bottleneck of iterative time-marching schemes. 
Third, an efficient measurement algorithm extracts statistical moments from the final quantum state using low-order polynomial approximations with a complexity polynomial in $\log N$. 
Together, these components establish a pathway for applying quantum computing to nonlinear reacting flows, from the formulation of the governing equations to the extraction of physically meaningful results.

The framework was validated in two stages. 
The \textit{a priori} tests on beta distributions confirmed that the measurement algorithm accurately recovers the mean and variance across a wide range of PDF shapes, with the sixth-order approximation reproducing the ground truth with high fidelity. 
The complete framework was then demonstrated on a PSR model,  where the history state method successfully captured the full PDF evolution. 
The statistical moments extracted from the resulting quantum state converge rapidly with increasing polynomial order, achieving average relative errors of 0.03\% for the mean and  0.24\% for the variance at sixth order. 
These results demonstrate the proposed framework provides a quantum workflow from equation solving to physical information extraction. 

However, the current framework has several limitations. 
First, the validation is restricted to a spatially homogeneous system with a single composition variable, which avoids the full complexity of turbulent reacting flows.
Extension to multi-species systems will increase the dimensionality of the PDF transport equation, though the exponential scaling of the quantum state space is expected to mitigate this cost. 
Second, the substantial quantum resources required by the present algorithm exceed the capabilities of NISQ devices, positioning it as an algorithm for the fault-tolerant era~\cite{Katabarwa2024_Early}.
Nevertheless, the development and validation of such algorithms are essential to prepare for future hardware capabilities.

Future work will proceed in several directions. 
First, investigate resource-optimized QLSAs~\cite{Costa2022,Wang2024} to reduce the qubit and gate counts required for implementation. 
Second, extend the framework to multi-species systems and spatially inhomogeneous configurations, incorporating molecular diffusion models to address more realistic combustion problems.
Third, the integration of the proposed quantum framework with classical solvers in a hybrid approach may offer a practical pathway to leverage near-term quantum hardware for selected components of the simulation. 

\acknowledgement{CRediT authorship contribution statement} \addvspace{10pt}

{\bf JZ}: Conceptualization, Methodology, Software, Investigation, Data Curation, Visualization, and Writing - Original Draft. 
{\bf ZY}: Methodology, Software, Investigation, Data Curation, and Visualization. 
{\bf ZM}: Methodology, Formal analysis, and Writing - Review \& Editing. 
{\bf ZL}: Conceptualization, Formal analysis, Visualization, Writing - Review \& Editing, Supervision, and Funding acquisition.
{\bf YY}: Conceptualization, Writing - Review \& Editing, Supervision, and Funding acquisition.

\acknowledgement{Declaration of competing interest} \addvspace{10pt}


The authors declare that they have no known competing financial interests or personal relationships that could have appeared to influence the work reported in this paper.

\acknowledgement{Acknowledgments} \addvspace{10pt}

This work has been supported in part by the National Natural Science Foundation of China (Nos. 52306126, 12525201, 12432010, and 12588201), the Beijing Natural Science Foundation (No. F261001), and the National Key R\&D Program of China (No. 2023YFB4502600).

\footnotesize
\baselineskip 9pt

\clearpage
\thispagestyle{empty}
\bibliographystyle{proci}
\bibliography{QPDF}


\newpage

\small
\baselineskip 10pt


\end{document}